\documentstyle[preprint,aps]{revtex}

 \newcommand \be {\begin{equation}}
\newcommand \bea {\begin{eqnarray} \nonumber }
\newcommand \ee {\end{equation}}
\newcommand \eea {\end{eqnarray}}
 
 \newcommand \bi {\bibitem}
\newcommand \s {\sigma}
\newcommand \de {\delta}
\newcommand \De {\Delta}
\newcommand \g {\gamma}

\newcommand \la {\lambda}

 \newcommand \al {\alpha}

\begin{document}
\draft
\preprint{MA/UC3M/10/96}
\title{Closure of the Monte Carlo dynamical equations in the 
spherical Sherrington-Kirkpatrick model.}

\author{L. L. Bonilla(*), F. G. Padilla(*), G. Parisi(**) 
and F. Ritort(*)}
\address{(*) Departamento de Matem{\'a}ticas,\\
Universidad Carlos III, Butarque 15\\
Legan{\'e}s 28911, Madrid (Spain)\\
E-Mail: bonilla@ing.uc3m.es\\
E-Mail: padilla@dulcinea.uc3m.es\\
E-Mail: ritort@dulcinea.uc3m.es\
\\ (**) Dipartimento di Fisica,\\
Universit{\`a} di Roma I {\sl ``La  Sapienza''}
\\ INFN Sezione di Roma I \\ Piazzale
Aldo Moro, Roma 00187\\
E-Mail: parisi@vaxrom.roma1.infn.it}

\date{\today}
\maketitle

\begin{abstract}
We study the analytical solution of the Monte Carlo dynamics in the
spherical Sherrington-Kirkpatrick model using the technique of the
generating function. Explicit solutions for one-time observables (like
the energy) and two-time observables (like the correlation and
response function) are obtained. We show that the crucial quantity
which governs the dynamics is the acceptance rate. At
zero temperature, an adiabatic approximation reveals that the
relaxational behavior of the model corresponds to that of a single
harmonic oscillator with an effective renormalized mass.
\end{abstract} 

\vfill

\vfill

\narrowtext
\section{Introduction}

The off-equilibrium features of glassy systems constitute a very
interesting field of research. A great deal of attention has been paid
to the study of the dynamics of exactly solvable models of disordered 
systems where much relevant information can be gathered even in
the mean-field case. Numerical simulations of short-ranged and
long-ranged models reveal a similar nature of the off-equilibrum dynamics
\cite{Mattsson,Rieger,CuKuRi,KiSaScRi}. The outstanding phenomenon which
characterizes the off-equilibrium regime is the presence of {\em aging},
which has been experimentally observed in a large class of physical
systems (for instance, in spin glasses \cite{suecos} as well as in real
glasses \cite{Struik}). The distinctive feature of the aging
phenomenon is that the response of the system to an external
perturbation depends strongly on the time elapsed
since the perturbation was applied. A phenomenological approach was
proposed by Bouchaud \cite{Bou} where the slow
dynamics originates from the presence of large number of traps in phase
space with a very broad distribution of lifetimes.

Much interest has been also devoted to the study of exactly solvable
mean-field spin-glass models where information can be obtained from the
solution of the dynamical equations. Of particular interest is the study
of disordered models where replica symmetry is broken in the
low-temperature phase. In this case, several results have been obtained
in the study of $p$-spin models\cite{KiTh,CHS,CuKu} as well as in the
study of a particle in a random potential \cite{FrMe,CuDo,Ho}. From
these studies a very interesting connection \cite{BoCuKuMe} between mean-field 
glassy dynamical equations and the Mode Coupling Theory of glasses 
\cite{MCT,KiTh2} has emerged.  In
this theoretical framework one derives a closed system of integrodifferential 
equations for the two-time correlation and response functions. Whereas all
the information about the dynamics is contained in this system of equations,
it is very difficult ascertain the long time evolution of
one-time observables like the energy or the magnetization.  Despite 
several advances in this direction (see for instance, the time series
expansion carried out by Franz, Marinari and Parisi in case of the $p=3$
spherical spin glass model \cite{FrMaPa}) this is still an open problem.
Recently a completely different approach to the dynamics of glassy systems
has been proposed by Coolen and Sherrington \cite{CS}. They
introduce a closure assumption (the equipartitioning hypothesis) which yields 
evolution equations for the one-time observables. In comparison with the results 
of simulations they obtain fairly good results in the short time regime. 
However it is still unclear whether the
equipartitioning approximation is applicable for long times and how to proceed
in case it this is the case \cite{Jap,Parisi}.  

On the issue of spin-glass dynamics, most of the existing works study 
Langevin equations. Meanwhile Monte Carlo dynamics has received less
attention notwithstanding its great importance in numerical simulations,
\cite{Rieger2}.  In particular, given that a major part of the
numerical work in spin-glasses uses the Monte Carlo algorithm, 
we think that it is important to study the Monte Carlo dynamics in itself.
Notice that Monte Carlo dynamics introduces the concept of
acceptance rate which has no meaning for the Langevin dynamics. Interestingly,
we shall see that the nature of the off-equilibrium regime strongly
depends on the behavior of the acceptance rate. To shed some light on
these issues we have considered the spherical Sherrington-Kirkpatrick
(SK) model \cite{KTJ}. While this model lacks replica symmetry breaking in 
the low temperature phase, it can be analytically solved.

In a letter (hereafter referred as I) we introduced a method to
solve the Monte Carlo dynamics of the spherical SK model \cite{nostre}. We
derived generalized dynamical equations for one-time quantities and a closed 
integrodifferential problem for their generating function. 
The same method can be successfully applied to models with Langevin dynamics.
The purpose of this work is to
present the method of the generating function, give a detailed account
of the computations announced in I, and discuss some other topics
such as the properties of two-time correlation and
response functions and the long time behavior of dynamical quantities.

The paper is organized as follows. In the second section we introduce the
spherical SK model and explain how to solve its Langevin dynamics by the
method of the generating function. While the results of this section are well
known \cite{CiPa,CugDean}, their derivation by means of the generating 
function serves as a clear introduction of this method.  Section three contains 
the main result of this paper, i.e. the  solution of the corresponding equations
for the Monte Carlo dynamics. Section four
analyzes the long-time behavior of dynamical quantities at finite and zero
temperature and the conditions under which the results of the Langevin 
dynamics are recovered. The last section contains our conclusions while the 
Appendices are devoted to different technical matters.

\section{The Langevin approach: a reminder}

The spherical Sherrington-Kirkpatrick spin glass model is defined by the
Hamiltonian,

\be
{\cal H}\lbrace\s\rbrace=-\sum_{i<j}\,J_{ij}\s_i\s_j
\label{eq21}
\ee

\noindent
where the indices $i,j$ run from $1$ to $N$ ($N$ is the number of
sites) and the spins $\s_i$ satisfy the spherical global 
constraint
\be
\sum_{i=1}^N\,\s_i^2=N.
\label{eq22}
\ee

The interactions $J_{ij}$ are Gaussian distributed with zero mean and
$1/N$ variance. The statics of this model reveals the existence of a 
thermodynamic second-order phase transition at $T=1$ \cite{KTJ} with a replica
symmetric low-temperature phase.

We will revisit the Langevin dynamics of the spherical SK model in order
to introduce the technique of the generating function. The Langevin 
equation is 

\be 
\frac{\partial \sigma_i}{\partial t}=-\frac{\partial {\cal
H}}{\partial \sigma_i}+\mu(t)\sigma_i+\eta_i(t)
\label{eq23}
\ee

\noindent
with $\eta_i(t)$ a Gaussian white noise
$\overline{\eta_i(t)\eta_j(t')}=2T\delta_{i,j}\delta(t-t')$,
($\overline{\cdot\cdot\cdot}$ stands for average over the noise); $\mu(t)$ 
is a time dependent Lagrange multiplier which ensures that the spherical
constraint eq.(\ref{eq22}) is satisfied at all times and ${\cal H}$ is
the Hamiltonian defined by eq.(\ref{eq21}).We can rewrite the previous
equation as a single mode problem,

\be
\frac{\partial \sigma_{\la}}{\partial
t}=\la\sigma_{\la}+\mu(t)\sigma_{\la}
+\eta_{\la}(t)
\label{eq24}
\ee

Here $\sigma_{\la}$ and $\eta_{\la}$ are the projections of the
configuration $\lbrace\sigma_i\rbrace$ and the original noise
$\lbrace\eta_i\rbrace$ on the basis of eigenvectors which diagonalize
the interaction matrix $J_{ij}$. The $\lbrace\sigma_{\la}\rbrace$ satisfy
the spherical constraint $\sum_{\la=1}^N\,\s_{\la}^2=N$. The transformation
matrix which diagonalizes the $J_{ij}$ is an orthogonal matrix.  Hence, the 
components $\eta_{\la}$ generate a white noise of the type,

\be
\overline{\eta_{\la}(t)\eta_{\la'}(t')}=2
T\delta_{{\la},{\la'}}\delta(t-t')~~~~~~~~~~.
\label{eqa21}
\ee

The eigenvalues 
$\la$ of a random Gaussian symmetric matrix are distributed according to 
the Wigner semicircular law $w(\la)$ \cite{WIGNER},

\be
w(\la)=\frac{\sqrt{4-\la^2}}{2\pi} 
\label{wigner}
\ee

\subsection{Generating function for the one-time quantities}

Our purpose is to describe the time-evolution of  certain one-time functions
of the solutions of eq.(\ref{eq24}). We define the set of moments,

\be
h_k=\frac{1}{N}\sum_{(i,j)} \overline{\s_i(J^k)_{ij}\s_j}=\frac{1}{N}
\sum_{\la}\la^k \overline{\s_{\la}^2}
\label{eqa22}
\ee

\noindent
Notice that  $h_0=1$ (spherical constraint) and $h_1=-2E$ where $E$ is the
energy. Using the result $\lim_{t\to
t'}\overline{\eta_{\lambda}(t')\sigma_{\lambda}(t)}=2T$ 
we get the equation

\be
\frac{\partial h_k}{\partial t} = 2h_{k+1}+2\mu h_k+ 2 T \ll\la^k\gg 
\label{eqa23}
\ee

\noindent
where 

\be
\ll f(\la)\gg =\int_{-2}^{2} f(\la)\,w(\la)\,d\la
\label{eqa23b}
\ee

\noindent
and $w(\la)$ is given by eq.(\ref{wigner}). In particular,
eq.(\ref{eqa23}) for $k=0$ gives the Lagrange multiplier as a function
of the energy and temperature $\mu=2E-T$. In order to close these
equations we define the generating function,

\be g(x,t)=\frac{1}{N}\sum_{(i,j)}\s_i\,(e^{xJ})_{ij}\,\s_j=
\frac{1}{N}\sum_{\la} e^{\la
x}\s_{\la}^2(t)=\sum_{k=0}^{\infty}\frac{x^k}{k!}h_k(t).
\label{eqa24}
\ee

\noindent
This function yields all the moments $h_k=\Bigl (\frac{\partial^k
g(x,t)}{\partial x^k}\Bigr )_{x=0}$.

We now want to formulate a problem for $g(x,t)$ (equation, initial, boundary 
and subsidiary conditions) which has a unique solution. Then we will either
be able to solve it (and thus determine the $h_k's$ exactly) or to infer the
long-time behavior of the moments. By using Equations (\ref{eqa22}), (\ref{eqa23})
and (\ref{eqa24}), we find that $g(x,t)$ satisfies the following differential
equation

\begin{eqnarray}
\frac{\partial g(x,t)}{\partial t} =2\frac{\partial g(x,t)}{\partial
x}+2 \mu(t) g(x,t)+2T\ll\exp(x\la)\gg ,\label{eqa25}\\
\mu(t) = - {\partial g\over \partial x}(0,t) - T.
\label{eqa25a}
\end{eqnarray}
These equations have to be solved with the boundary condition $g(0,t) = 1$
(the spherical constraint) and an appropriate initial condition $g(x,0)=g_0(x)$
(defined by the initial configuration $\lbrace\s_{\la}(t=0)\rbrace$).
Besides taking the continuum limit in the definition (\ref{eqa24}), we see
that 
\be
g(x,t) = \int_{-2}^2 d\lambda\, w(\lambda)\, e^{x\lambda}\, \hat{g}(\lambda,t),
\ee
where the spectral transform $\hat{g}(\lambda,t)\geq 0$. The latter condition
is kept by the dynamics if it holds initially. However its use is crucial
to distinguish the physically meaningful stationary solutions from spurious ones. 
We can solve (\ref{eqa25}) by the method of characteristics assuming that 
$\mu(t)$ is given. The result is

\be
g(x,t)=g_0(x+2t)e^{2\int_0^t\,\mu(t')dt'}\,+\,2T\int_0^t\,dt'
\ll e^{(x+2(t-t'))\la}\gg \,e^{2\int_0^{t'}\,\mu(t'')dt''}
\label{eqa25b}
\ee

\noindent
Inserting (\ref{eqa25b}) in the spherical constraint $g(0,t)=1$, it is possible 
to derive an integral equation which can be solved by means of the Laplace 
transform \cite{CugDean}.

It is easy to check that (\ref{eqa28})  is a 
stationary solution of eq.(\ref{eqa25}) for $T > T_{c}=1$.

\be
g^{eq}(x)=-\ll\frac{e^{x\la}}{\beta(\la+\mu^{eq})}\gg
\label{eqa28}
\ee

In this case the moments $h_k$ can be easily computed
\cite{KTJ}

\be
h_k^{eq}=-\ll\frac{\la^k}{\beta(\la+\mu^{eq})}\gg,
\label{eqa26}
\ee

\noindent
and the Lagrange multiplier $\mu^{eq}$ is given by

\be
1=-\ll\frac{1}{\beta(\la+\mu^{eq})}\gg 
\label{eqa27}
\ee

\noindent
For $T \leq 1$, $\mu^{eq}=-2$, the stationary solution is given by

\be
g^{eq}(x)=(1-\frac{1}{\beta}) e^{2 x}
-\ll\frac{e^{x\la}}{\beta(\la-2)}\gg,
\label{eqa28b}
\ee

\noindent
and the moments $h_k$ are 
\be
h_k^{eq}=(1-\frac{1}{\beta}) 2^k -\ll\frac{\la^k}{\beta(\la-2)}\gg,
\label{eqa26b}
\ee

\noindent
Substituting eqs.(\ref{eqa28} and \ref{eqa28b}) in eq.(\ref{eqa25}) one
can easily check that the right hand side of eq.(\ref{eqa25}) is 
identically zero.

\subsection{The correlation and response function }

In order to find closed expressions for the correlation function we
define the following set of two-times moments,

\be
C_k(t',t)=\frac{1}{N}\,\sum_{(i,j)} \overline{\s_i(t')(J^k)_{ij}\s_j(t)}=
\frac{1}{N}\,\sum_{\la}\la^k\overline{\s_{\la}(t')\,\s_{\la}(t)}
\label{eqb21}
\ee

Note that in this notation the usual two-times correlation function is
given by $C_0(t',t)$. The equation of motion for the $C_k(t',t)$ reads,

\be
\frac{\partial C_k(t',t)}{\partial t}=C_{k+1}(t',t) + \mu(t)\,C_k(t',t) 
\label{eqb22}
\ee

\noindent
where we have used the result 
$\overline{\eta_{\lambda}(t')\sigma_{\lambda}(t)}=0$ for
$t'<t$. Previous equation has to be solved with the initial condition 
$C_k(t',t')=h_k(t')$ where the $h_k(t)$ are the one-time moments
previously obtained from the generating function $g(x,t)$. 

We define the following generating function,

\be
K(x,t',t)=\frac{1}{N}\sum_{(i,j)}\overline{\s_i(t')\,(e^{xJ})_{ij}\,\s_j(t)}=
\frac{1}{N}\sum_{\la} e^{\la
x}\overline{\s_{\la}(t')\s_{\la}(t)}=\sum_{k=0}^{\infty}\frac{x^k}{k!}C_k(t',t).
\label{eqb23}
\ee

The generating function $K(x,t',t)$ yields the generalized two-times
moments $C_k(t',t)=\Bigl (\frac{\partial^k
K(x,t',t)}{\partial x^k}\Bigr )_{x=0}$ and satisfies the following
homogeneous partial differential equation,

\be
\frac{\partial K(x,t',t)}{\partial t} =\frac{\partial K(x,t',t)}{\partial
x}+\mu(t) K(x,t',t)
\label{eqb24}
\ee

\noindent
with the initial condition $K(x,t',t')=g(x,t')$ and $g(x,t')$ is given in
eq.(\ref{eqa25b}).

A similar method is applied to the response function. We define the set
of two-times moments,

\be
G_k(t',t)=\frac{1}{N}\,\sum_{(i,j)}(J^k)_{ij}\overline{\frac{\partial\s_j(t)}
{\partial \eta_i(t')}}=\frac{1}{N}\,
\sum_{\la}\la^k \overline{\frac{\partial \s_{\la}(t)}{\partial \eta_{\la}(t')}}
\label{eqb25}
\ee

\noindent
where $t'<t$. In this notation the usual response function is given by
$G_0(t',t)$. 
We construct the generating function  

\bea
\Gamma(x,t',t)=\frac{1}{N}\sum_{(i,j)}\,(e^{xJ})_{ij}\,\overline{
\frac{\partial\s_j(t)} {\partial\eta_i(t')}}=
\frac{1}{N}\sum_{\la} e^{\la x}\overline{\frac{\partial\s_{\la}(t)}{\eta_{\la}(t')}}\\
=\sum_{k=0}^{\infty} \frac{x^k}{k!}G_k(t',t).
\label{eqb26}
\eea

The generating function $\Gamma(x,t',t)$ yields the generalized
two-times moments $G_k(t',t)=\Bigl (\frac{\partial^k \Gamma
(x,t',t)}{\partial x^k}\Bigr )_{x=0}$. With the usual regularization of
the response function at equal times the dynamical equation for the
$\Gamma(x,t',t)$ reads,

\be
\frac{\partial \Gamma(x,t',t)}{\partial t}=\frac{\partial 
\Gamma(x,t',t)}{\partial x}+\mu(t)\Gamma(x,t',t)\,+\,
\delta(t-t')\ll\exp(x\la)\gg 
\label{eqb27}
\ee

To solve this equation we need to impose the causality condition
$\Gamma(x,t',t)=0$ if $t<t'$.
The solution to equations (\ref{eqb24}) and (\ref{eqb27}) can be easily
found. One gets the results,

\bea
K(x,t',t)=g(x+t-t',t')\,e^{\int_{t'}^{t}\mu(t'')dt''}\\
\Gamma(x,t',t)=\ll e^{(x+t-t')\la}\gg \,e^{\int_{t'}^t\mu(t'')dt''}
\theta(t-t')
\label{eqb28}
\eea

From these generating functions we can extract the usual two-time
correlation and response function, $C_0(t',t)=K(0,t',t)$ and
$G_0(t',t)=\Gamma(0,t',t)$. At equilibrium we can substitute the
solution (\ref{eqa28}) in (\ref{eqb27}) obtaining,

\bea
K(x,t',t)=K^{eq}(x,t-t')=g^{eq}(x+t-t')e^{\mu^{eq}(t-t')}\\
\Gamma(x,t',t)=\Gamma^{eq}(x,t-t')=\ll e^{(x+t-t')\la}\gg \,e^{\mu^{eq}(t-t')} 
\theta(t-t')
\label{eqb29}
\eea

Both functions are translationally invariant (i.e. depend only on the
difference of times) and the fluctuation
dissipation theorem for the generating functions also holds, 

\be
\Gamma^{eq}(x,t-t')=\beta\frac{\partial K^{eq}(x,t-t')}{\partial t'}
\label{eqb30}
\ee

The off-equilibrium behavior of the quantities $C_0(t',t), G_0(t',t)$
has been already studied in the literature (see \cite{CiPa,CugDean}).

\section{The Monte Carlo approach}

We consider the Monte Carlo (MC) dynamics with the Metropolis
algorithm. The idea behind the MC approach is to postulate a dynamics in
which a new configuration is proposed and accepted with a certain
probability. The dynamics is ergodic at finite temperature and satisfies
detailed balance. Following (I) we will consider a
particularly simple motion which makes the dynamics exactly soluble: 
take the configuration $\lbrace\s_i\rbrace$ at time $t$ and perform a
small random rotation from that configuration to a new one
$\lbrace\tau_i\rbrace$ where

\be
\tau_i=\s_i+\frac{r_i}{\sqrt{N}}
\label{eq31}
\ee

\noindent
and the $r_i$ are random numbers
extracted from a Gaussian distribution $p(r)$ of finite 
\mbox{variance $\rho$,}
\be
p(r_i)=\frac{1}{\sqrt{2\pi\rho^2}}\exp(-\frac{r_i^2}{2\rho^2})~~~.
\label{eq32}
\ee

Let us denote by $\Delta E$ the change of energy $\Delta
E=E\lbrace\tau\rbrace-E\lbrace\s\rbrace$.  According to the Metropolis
algorithm we accept the new configuration with probability 1 if $\Delta
E< 0$ and with probability $exp(-\beta\Delta E)$ if $\Delta E> 0$ where
$\beta=\frac{1}{T}$ is the inverse of the temperature $T$.

\subsection{The probability distribution $P(\Delta E)$}

As before in the Langevin formulation, we want to obtain one and
two time moments. It is useful to work in the basis for which the 
interaction matrix $J_{ij}$ is diagonal. The energy then reads,

\be
E\lbrace\s_{\la}\rbrace=-\frac{1}{2}\sum_{\la}\la\,\s_{\la}^2
\label{eqa31}
\ee 
\noindent
where the $\s_{\la}$ are distributed according to eq.(\ref{wigner}). 
In this basis, the spin configuration still corresponds to a small 
random rotation, hence $\s_{\la}\to \s_{\la}+ r_{\la}/\sqrt{N}$ where 
the new random numbers $r_{\la}$ are extracted from the same Gaussian 
distribution eq.(\ref{eq32}).

The basic object we want to compute is the probability $P(\Delta E)$ of
having a given variation $\De E$ of the energy $E$. This is a quantity
which gives the average number of accepted changes.  The
variation $\De E^*$ of the energy $E$ in a Monte Carlo (MC) step
is 
\bea \De
E^*=-\frac{1}{\sqrt{N}}\sum_{\la}\la\,\s_{\la}r_{\la}-
\frac{1}{2N}\sum_{\la}\,\la\,r_{\la}^2\\
\label{eqa33}
\eea

\noindent
while the quantity $h_0=\frac{1}{N}\,\sum_{\la}\s_{\la}^2$
is changed by the following amount,

\be 
\De h_0^*=\frac{2}{\sqrt{N}}\sum_{\la}\s_{\la}\,r_{\la}+
\frac{1}{N}\sum_{\la}\,r_{\la}^2.
\label{eqa34}
\ee

The probability $P(\Delta E)$ of having a change of energy is

\be
P(\Delta E)=\int\de(\De E-\De E^*)
\de(\De h_0^*)\prod_{\la} \Bigr( p(r_{\la})dr_{\la}\Bigl)
\label{eqa35}
\ee
\noindent
where the last delta function in the integrand accounts for the
spherical constraint and the variation $\De E^*$ is given in
eq.(\ref{eqa33}). 

Using the integral representation for the delta function 
\be
\de(x)=\frac{1}{2\pi}\int_{-\infty}^{\infty}e^{i\alpha x}d\alpha
\label{eqa36}
\ee

\noindent
and substituting in (\ref{eqa35}) we get
\bea
P(\Delta E)=\int d\mu\, d\eta\, 
\exp\Bigl (i\mu\De E -\\
\frac{\rho^2}{2N}\sum_{\la}\frac{\s_{\la}^2\g_{\la}^2}
{(1-\frac{i\g_{\la}^2
\rho^2}{N})}-\frac{1}{2}\sum_{\la}\log(1-\frac{i\g_{\la}^2
\rho^2}{N})\Bigr )
\label{eqa37}
\eea

\noindent
where $\g_{\la}=\mu \la + 2\eta$. After expanding the
logarithm and retaining the first $1/N$ correction we get (after some
 manipulations)
\be
P(\De E)=\frac{1}{\sqrt{2\pi\rho^2 B_1}}\,\exp\Bigl (-\frac{(\De E+\rho^2 E)^2}
{2\rho^2B_1}\Bigr )
\label{eqa38}
\ee
\noindent
with
\be
B_1=h_{2}-4E^2;~~~(h_0=1);~~~h_2=\frac{1}{N}\sum_{\la}\,\la^2\,\s_{\la}^2;
\label{eqa39}
\ee

The equation for the energy is obtained by considering the average
variation of energy in an MC step. In this case one 
MC step corresponds to $N$ elementary moves. In the thermodynamic
limit we can write the continuous equations,

\be \frac{\partial E}{\partial t} = \overline{\De
E}=\int_{-\infty}^{0}(\De) E\,P(\De E)\,d(\De E)+ \int_{0}^{\infty}(\De
E)\,\exp(-\beta \De E)\,P(\De E)\,d(\De E)~~~~~.
\label{eqa300}
\ee

We can check that equilibrium is a stationary solution of the Monte
Carlo dynamics. Using standard static calculations \cite{MePaVi,KTJ} one can
show that in equilibrium $B_1= h_2-4E^2=-2ET$. In this case a
straightforward computation shows that detailed balance is
fulfilled. This means that, for a given value of $\De E$, the first
integral appearing in the r.h.s of eq.(\ref{eqa300}) cancels the
contribution of the second integral in the r.h.s of eq.(\ref{eqa300})
for the same value of $\De E$. In other words,

\be
P(-\De E)=\exp(-\beta \De E)P(\De E)
\label{det_bal}
\ee

The equation for the energy reads,

\be \frac{\partial E}{\partial t} = -\frac{a(t)}{2}h_2(t)+b(t)E(t)
\label{eneTf}
\ee

\noindent
where the coefficients $a(t)$ and $b(t)$ are given by

\bea
a(t)=\rho^2\beta\,e^{\frac{\rho^2\beta}{2}(\beta
B_1+2E)}\,Erf(\rho\beta\sqrt{\frac{2}{B_1}}-\alpha)\\
b(t)=-\frac{1}{2}(\rho^2 Erf(\alpha)+(4E-T)a(t))
\label{coefene}
\eea

\noindent
and $Erf(x)$ is the complementary error function defined as
$Erf(x)=\frac{2}{\sqrt{\pi}}
\int_{x}^{\infty}\,dx\exp(-x^2)$ and the parameter $\alpha$ is
given by,
\be
\alpha=-\frac{\rho E}{\sqrt{2B_1}}
\label{eqa303}
\ee

Note that the quantities E, $B_1$, and $\alpha$ depend on time.
Also one can compute the acceptance rate as a function of time, which is
the probability of accepting a certain change of the configuration,
\be
A(t)=\int_{-\infty}^{0}\,P(\De E)\,d(\De E)+
\int_{0}^{\infty}\exp(-\beta \De E)\,P(\De E)\,d(\De E)~~~~~.
\label{eqa301}
\ee

A straightforward computation shows,

\be
A(t)=\frac{Erf(\alpha)}{2}+\frac{1}{2}e^{\frac{\rho^2\beta}{2}(\beta
B_1+2E)}\,Erf(\rho\beta\sqrt{\frac{2}{B_1}}-\alpha)
\label{eqa302}
\ee

In order to obtain the time evolution of the acceptance rate and to
solve the eq.(\ref{eneTf}) of the time evolution of the energy we need
to know the energy $E$ and $h_2$ at time $t$. Unfortunately, one can see
that the time evolution equation for $h_2(t)$ involves $h_3(t)$ and so
on. This hierarchy of moments can be closed introducing a generating
function \cite{nostre} like has been done in the Langevin dynamics. This
is the purpose of the next section.

\subsection{Generating function for the one-time quantities}

To close the equations of motion we consider the set of 
moments defined in eq.(\ref{eqa22}). The basic object to compute is the
joint probability distribution $P(\Delta h_k,\Delta E)$. This quantity
can be written as,

\be
P(\Delta h_k,\Delta E)=\int\de(\De h_k-\De h_k^*) \de(\De E-\De E^*)
\de(\De h_0)\prod_{\la} \Bigr( p(r_{\la})dr_{\la}\Bigl)
\label{eqb31}
\ee

\noindent
where the last delta function in the integrand accounts for the
spherical constraint and the variation $\De E^*$ is given in
eq.(\ref{eqa33}) while the variation $\De h_k^*$ is given by, 

\be
\De h_k^*=\frac{2}{\sqrt{N}}\sum_{\la}\la^k\,\s_{\la}\,r_{\la}+
\frac{1}{N}\sum_{\la}\,\la^k\,r_{\la}^2.
\label{eqb32}
\ee

Following the same technical steps as in the derivation of $P(\De E)$
we obtain the following result,

\be
P(\Delta h_k,\Delta E)=P(\De E)\,P(\Delta h_k|\Delta E)
\label{eqb33}
\ee where $P(\De E)$ is the probability distribution eq(\ref{eqa38}) and
$P(\De h_k|\De E)$ is the conditional probability of $\De h_k$ given
$\De E$. The final expression for the conditional probability is,

\be
P(\Delta h_k|\Delta E)=\frac{1}{\sqrt{8\pi\rho^2(C_k-(B_k^2/B_1))}}\,
\exp(-\frac{(\De h_k+\rho^2(h_k-\ll \la^k\gg )+2\frac{B_k}{B_1}(\De
E+\rho^2 E))^2}{8\rho^2(C_k-B_k^2/B_1)}\Bigr )
\label{eqb34}
\ee
\noindent
with $C_k=h_{2k}-h_{k}^2;\,B_k=h_{k+1}+2Eh_{k};\,(h_0=1;h_1=-2E);$ and
the average $\ll...\gg $ has been previously defined in eq.(\ref{eqa23b}).

In order to obtain the dynamical evolution of the moments $h_k$ we
have to compute its average variation in a MC step over the
accepted changes of configuration. In the
thermodynamic limit we can write the continuous equations,

\bea
\frac{\partial h_k}{\partial t}=\overline{\De h_k}=
\int_{-\infty}^{\infty}d(\De h_k)\,\De h_k\\ 
\Bigl (\int_{-\infty}^{0}d(\De E)\,
P(\De h_k,\De E)+\int_{0}^{\infty}d(\De E)
\,\exp(-\beta\De E)\,P(\De h_k,\De E)\Bigr )
\label{eqb36}
\eea

The solution for a general integral of the previous type is shown in the
Appendix A. The following result is obtained,

\be
\frac{\partial h_k(t)}{\partial t}=a(t)h_{k+1}(t)+b(t)h_{k}(t)+c_k(t)
\label{eqb37}
\ee

\noindent
where the time dependent quantities $a(t)$ and $b(t)$ are given
in eq.(\ref{coefene}) and the coefficients $c_k(t)$ are defined by

\be
c_k(t)=(2E(t)a(t)-b(t))\ll\la^k\gg =\rho^2 A(t) \ll\la^k\gg.
\label{eqforc}
\ee

\noindent
Here $A(t)$ is the acceptance rate defined in eq.(\ref{eqa302}).  Note
that the rate variation of the moment $h_k$ depends linearly on the
moments $h_k$ and $h_{k+1}$, but the coefficients $a(t)$, $b(t)$
and $c(x,t)$ are nonlinear functions of $h_{1}=-\frac{1}{2} E$
and $B_1=h_2-h_1^2$ (second cumulant). It is thus reasonable to expect 
that the Monte Carlo dynamics is determined by the evolution of the first
two moments. By means of the moments generating function $g(x,t)$ of eq.
(\ref{eqa24}), we obtain from eq.  (\ref{eqb36}).

\be
\frac{\partial g(x,t)}{\partial t}=a(t)\frac{\partial g}
{\partial x}\,+\,b(t) g\,+\,c(x,t).
\label{eqb38}
\ee

\noindent
where the time dependent quantities $a, b$ are functions of the two first
moments $E(t)$ and $h_2(t)$ (whose relation to $g$ is indicated below)
 defined in (\ref{coefene}) and

\be
c(x,t)=(2E(t)a(t)-b(t))\ll e^{x\la}\gg =\rho^2 A(t)\ll e^{x\la}\gg ~.
\label{eqcx}
\ee

As in the case of Langevin dynamics, $g(x,t)$ is a solution of eq. 
(\ref{eqb38}) plus the following initial, boundary and subsidiary conditions:
\be
\begin{array}{lr}
g(x,0)=&g_0(x) \\
g(0,t)=&1 \\
\frac{\partial g}{\partial x}\mid_{x=0} =& -2 E(t)\\
\frac{\partial^2 g}{\partial x^2}\mid_{x=0}=&h_2(t)\\
\end{array}
\ee

The second condition is the spherical one and the third and fourth define 
the first and second moments of the set $h_k$.
This linear partial differential equation equation can be readily solved
with the method of characteristics. The general solution for a partial
differential equation of the previous type (\ref{eqb38}) is shown in the
Appendix B.

\subsection{The correlation and response function}

In order to find the dynamical equation for the set of correlation
functions eq.(\ref{eqb21}) we perform a similar computation as has been
done for the moments $h_k$.


The elementary move eq.(\ref{eq31}) at time $t$ induces a change $\De
E^*$ and $\De C_k^*$ in the energy and the $k$-moment of the correlation
function $C_k(t',t)$ (in what follows we take $t'<t$) defined in
eq.(\ref{eqb21}),

\bea
\De E^*=-\frac{1}{\sqrt{N}}\sum_{\la}\la\s_{\la}(t)r_{\la}(t)-
\frac{1}{2N}\sum_{\la}\,\la\,r_{\la}(t)^2\\
\De C_k^* = \frac{1}{\sqrt{N}}\sum_{\la}\la^k\s_{\la}(t')r_{\la}(t)
\label{eqc30}
\eea

The probability of  having a change of the energy $\Delta E$ 
and the correlation $\De C_k$ is given by the joint probability,

\bea
P(\Delta E,\De C_k)=\int\de(\De E-\De E^*) \de(\De C_k-\De C_k^*)
\de(\De h_0)\prod_{\la} \Bigr( p(r_{\la})dr_{\la}\Bigl)
\eea


Using the integral representation of the delta function, 
 retaining only the terms of order $\frac{1}{N}$ and performing all the
Gaussian integrals we find 
$P(\De E, \De C_k)= P(\Delta E) P(\De E| \De C_k)$ where, 
\bea
P(\De E| \De C_k)= \frac{1}{\sqrt{2 \pi \rho^2
(h'_{2k}-C_k^2 - \frac{D_k^2}{B_1})}}
\exp\Bigl (-\frac{(\De C_k+\frac{\rho^2}{2}C_k+
\frac{D_k}{B_1} (\De E -\frac{\rho^2}{2} h_1) )^2}
{2 \rho^2 (h'_{2k}-C_k^2 - \frac{D_k^2}{B_1}) }\Bigr )
\eea

\noindent
where $h'_{2k}=h_{2k}(t');\,\,D_k(t',t)=C_{k+1}(t',t)+2E(t)C_k(t',t)$ and
$P(\Delta E)$ is given by eq.(\ref{eqa38}).\\


To solve the equation of motion for the $C_k(t',t)$ we write its average
variation at time $t$ over the accepted changes of configuration,

\bea
\frac{\partial C_k(t',t)}{\partial t}=\overline{\De C_k(t',t)}=
\int_{-\infty}^{\infty}\De C_k\,d(\De C_k)\,\\ \Bigl (\int_{-\infty}^{0}
d(\De E)\,P(\De C_k,\De E)+\int_{0}^{\infty}d(\De E)\,
\exp(-\beta\De E)\,P(\De C_k,\De E)\Bigr )
\label{eqc32b}
\eea

Using the formulae of Appendix A we get the result,
\be
\frac{\partial C_k(t',t)}{\partial t} = \frac{a(t)}{2} C_{k+1}(t',t)
+\frac{b(t)}{2}C_k(t',t)
\label{eqc33}
\ee

\noindent
where $a(t), b(t)$ have been previously defined in 
eq.(\ref{coefene}).

Equation (\ref{eqc33}) is solved with the initial condition
$C_k(t',t')=h_k(t')$ (once the time evolution of the $h_k(t)$ has been
obtained solving the preceding hierarchy corresponding to the set of
moments $h_k$).

In order to close the previous hierarchy of equations we introduce the
generating function $K(x,t',t)$ of eq.(\ref{eqb23}) which yields the
generalized two-times moments $C_k(t',t)=\Bigl (\frac{\partial^k
K(x,t',t)}{\partial x^k}\Bigr )_{x=0}$.  The $K(x,t',t)$ satisfies the
following partial differential equation,

\be
\frac{\partial K}{\partial t}=\frac{a(t)}{2}\frac{\partial K}{\partial
x}\,+\,\frac{b(t)}{2} K
\label{eqc35}
\ee

\noindent
together with $K(x,t',t')=g(x,t')$. The solution can be found by
 the method of the characteristics.
A similar procedure can be used to obtain the generating function for
the moments of the response function eq.(\ref{eqb25}). In the Monte
Carlo dynamics the equivalent quantity is given by the moments,

\be
G_k(t',t)=\lim_{\De\to 0}\frac{1}{N\De}\sum_{\la}\la^k\,m_{\la}(t')\s_{\la}(t)
\label{eqc37}
\ee

\noindent
where $\De$ measures the intensity of an applied staggered field
$m_{\la}$.  This quantity measures the correlation between the spin
configuration at time $t$ and a small staggered magnetic field
$m_{\la}(t')$ applied at a previous time $t'$. Equation (\ref{eqc37}) is
the analogous of the correlation between the spin configuration at time
$t$ and the noise at time $t'$ in the Langevin approach. The staggered
magnetic field $m_{\la}$ is an uncorrelated annealed random field taken
from a Gaussian distribution of variance $\De$,

\be
P(m_{\la})=(2\pi\De^2)^{-\frac{1}{2}}\exp(-\frac{m_{\la}^2}{2\De^2})
\label{eqc38}
\ee

\noindent
at each elementary move in the MC dynamics. The calculation
of the response function proceeds in the following way: we compute the
probability distribution for the variation of the $G_k(t',t)$ when a
small staggered magnetic field $m_{\la}(t')$ is applied at an earlier
time $t'$ in an elementary move eq.(\ref{eq31}) at time $t$.  We
compute the joint probability distribution of the variation $\De
G_k(t',t)$ for an elementary move at time $t$,

\bea
P(\Delta E,\De G_k)=\int\de(\De E-\De E^*) \de(\De G_k-\De G_k^*)
\de(\De h_0)\\
\prod_{\la} \Bigr( p(r_{\la})P(m_{\la}(t'))dr_{\la}
dm_{\la}(t')\Bigl)
\label{eqc39}
\eea

\noindent
where $\De h_0$ is given in eq.(\ref{eqa34}) and the variations of
energy and $G_k$ are given by,

\bea
\De E^*=-\frac{1}{\sqrt{N}}\sum_{\la}\la\s_{\la}(t)r_{\la}(t)-
\frac{1}{2N}\sum_{\la}\,(\la\,r_{\la}(t)^2-m_{\la}(t')r_{\la}(t))\\
\De G_k^* = \frac{1}{\sqrt{N}}\sum_{\la}\la^k m_{\la}(t')r_{\la}(t)
\label{eqc301}
\eea

Finally, the perform the limit $\De$ tending to zero for the intensity of
the staggered magnetic field .
Performing similar computations as in the correlation function case we
obtain for the average variation $\frac{\partial G_k(t',t)}{\partial t}$
(in the region of times $t'<t$) the equation (\ref{eqc33}) substituting
$C_k$ by $G_k$. Note that the difference between the hierarchy of
equations associated to the correlation and the response function lies
in the region of times $t'>t$ where $G_k(t',t)=0$ while
$C_k(t',t)=C_k(t,t')$.

The partial differential equation
associated to the generating function eq.(\ref{eqb26}) is in this case

\be
\frac{\partial \Gamma}{\partial t}=a(t)\frac{\partial
\Gamma}{\partial x}\,+\,b(t) \Gamma\,+\,\ll\exp(x\la)\gg  \de(t-t')
\label{eqc302}
\ee

The partial differential equations (\ref{eqc35}) and (\ref{eqc302}) can
be also readily solved with the method of the characteristics (see
Appendix B) as was done for in case of the one time quantities. Note
that the equations for the generating functions $g,K,\Gamma$
(eqs.(\ref{eqb38}, \ref{eqc35}, \ref{eqc302})) are formally the same as in
the Langevin approach eqs.(\ref{eqa25},\,\ref{eqb24},\,\ref{eqb27}) with the
time-dependent parameters $a(t),b(t),c(x,t)$ given by,

\be
a_{LANG}(t)=2;~~~~~~~b_{LANG}(t)=2(2E-T);
~~~~~~~c_{LANG}(x,t)=2T\ll \exp(x\la)\gg ;
\label{eqc303}
\ee

The main difference between the MC dynamics and the Langevin dynamics
relies on the simpler time dependence of the coefficients $a,b,c$ in the
last case. This makes the large-time behavior of the Langevin dynamics
exactly soluble (see \cite{CiPa,CugDean}) while this is a very
complicated task in the MC case.

\section{Analysis of the Monte Carlo dynamical equations}

In this section we proceed to solve the resulting dynamical equations
for the MC dynamics. First we analyze the equilibrium dynamics
showing that it coincides with the Langevin dynamics by an appropriate
rescaling of time. Then we study the off-equilibrium behavior
contained in the MC dynamics at finite and zero
temperature. Since it is also our purpose to test the correctness of the
solution of our equations we will compare the theory with real Monte
Carlo numerical simulations. Moreover we will compare the resulting
dynamics with that expected in the Langevin case.

\subsection{Equilibrium Monte Carlo dynamics}

Dynamical equations for one-time and two-time quantities can be readily
solved at equilibrium. Now the observables $h_{k}$
are independent of time and the two-time functions $C_k(t',t)$,
$G_k(t',t)$ only depend on the differences of time
$t-t'$. It is easy to check
that the coefficients $a(t), b(t), c(x,t)$ of 
eqs.(\ref{coefene},\ref{eqcx}) are time-independent and given by,

\be
\begin{array}{lr}
a^{eq}=&\rho^2\beta Erf(\alpha^{eq})\\
b^{eq}=&\rho^2\beta (2E-T) Erf(\alpha^{eq})\\
c^{eq}(x)=&\rho^2\ll\exp(x\la)\gg  Erf(\alpha^{eq})
\end{array}
\label{eqeq} 
\ee

with

\be
\alpha^{eq}=-\frac{\rho E^{eq}}{\sqrt{2B_1^{eq}}}=
\rho\sqrt{\frac{(2\beta-1)}{8}}
\label{alpha}
\ee

This coefficients are the same as for the Langevin dynamics except for a
rescaling of time $t\to t'=\frac{\rho^2\beta
Erf(\alpha^{eq})}{2} t$. Also one can show (using
eq.(\ref{eqa302})) that the acceptance rate is given by,

\be
A^{eq}=Erf(\alpha^{eq})
\label{Aeq}
\ee

Using the previous rescaling of time the equilibrium MC dynamical
equations coincide with the Langevin ones\cite{footnote}.

\subsection{Monte Carlo simulations}

In order to check our analytical results for the MC dynamics we
have performed some MC simulations for finite sizes. 
To simulate enough large sizes we worked in the basis of
eigenvectors $\s_{\la}$. In this way all the information about the
quenched disorder is fully contained in the spectrum of eigenvalues
$\la$, which occupies much less memory than the full interaction
Gaussian matrix $J_{ij}$. The set eigenvalues $\la$ is chosen according
to the semicircular law eq.(\ref{wigner}). Typically we start
from a random initial configuration $\s_{\la}=\pm 1$ which fulfills the
spherical constraint. Then we perform a small rotation of this
configuration $\{\s_{\la}\}\to \{\s_{\la}+\frac{r_{\la}}{\sqrt{N}}\}$
(eq.(\ref{eq31})) where the $r_{\la}$ are random numbers extracted from
the Gaussian distribution eq.(\ref{eq32}). To keep invariant the
spherical constraint we normalize the length of the vector
$\lbrace\s_{\la}\rbrace$ in order to make it of length one.  The
resulting change of energy eq.(\ref{eqa31}) is computed and accepted
with probability $Min(1,\exp(-\beta\De E))$. We repeat this process for
the new configuration and so on. A MC step corresponds to $N$
rotations. Because an elementary move involves a global change of the
configuration $\lbrace\s_{\la}\rbrace$, this algorithm is $N$ times
slower than a usual MC algorithm with only local changes. We
are able to simulate relatively large sizes in the range $N=500-2000$
in a reasonable amount of computer time. 

\subsection{Finite-temperature dynamics}

In case of finite temperature it is relatively easy to solve MC
equations in the large-time limit. All the information on the dynamics
is contained in the time evolution of the coefficients $a(t), b(t),
c(x,t)$ which monotonically converge to their equilibrium values. 
then we expect the dynamics converges to the Langevin dynamics in the 
large-time limit except by a rescaling of the time

\be
t\to t'=\frac{\rho^2\beta Erf(\alpha^{eq})}{2} t
\label{renorm}
\ee

At finite temperature and according to the magnitude of the parameter
$\rho$ we distinguish two different regimes depending if the acceptance
rate is large or small. This two limits differ in the magnitude of the
time scale above which Langevin behavior is recovered.


\subsubsection{The case $\rho < 1$}

In this case $Erf(\alpha)\sim O(1)$ since $\alpha$ is small, i.e. the
acceptance is always large. 

Note that in the particular limit $\rho$ tending to zero the Langevin 
dynamics is recovered with a rescaling of time

\be
t\to t'=\frac{\rho^2\beta}{2} t
\label{renorm2}
\ee

In figure 1 we show the decay of the energy for different temperature
for a given value of $\rho=0.1$ by numerically solving the
eq.(\ref{Apb3}). The theoretical prediction is compared with real Monte
Carlo simulation results for $N=500$ spins. In the figure we also show
the asymptotic large time-behavior in the Langevin case
$E(t)-E^{eq}\sim\frac{3}{8}t^{-1}$ (See \cite{CugDean}) with the
rescaling of time eq.(\ref{renorm}).

Results for the correlation function $C_0(t_w,t_w+t)=K(0,t_w,t_w+t)$ are
shown in figure 2 for different values of $t_w=1,10,100,1000$ at the
temperature $T=0.4$ with $\rho=0.1$ obtained by numerically solving
eq.(\ref{Apb4}).  Again we compare the solution with real MC
simulations for $N=500$ spins. In the figure we also show the asymptotic
large-time behavior in the Langevin case  $C(t_w,t_w+t)\sim
t^{-\frac{3}{4}}$ (See \cite{CugDean}).

Note that in the dynamical regime shown in figure 2 there is no evidence
of plateau or $\beta$-relaxation process in the correlation
function. This process is a signature of the existence of a short-time
regime (FDT regime) where the system is locally in equilibrium and the
fluctuation-dissipation theorem applies (i.e. the $C_0(t_w,t_w+t)\simeq
C_0(t)$). The reason why the $\beta$-relaxation process is not seen in
the figure 2 is that the values of $t_w$ shown there are too small. In
the Langevin dynamics this regime usually appears for large values of $t_w$
which become much larger in the MC case since we have to rescale the
time by the parameter eq.(\ref{renorm}) (the constant we have to
rescale the time in figures 2 is approximately 87, i.e. 87 MC steps
correspond to one unit of time in the Langevin dynamics) . In this
regime $\rho\ll 1$, because $Erf(\alpha)$ is finite, we expect the $t_w$
necessary to observe the plateau to scale like $\rho^{-2}$ times the
needed time in the Langevin case which requires quite large values of
$t_w$ in the MC case. This off-equilibrium regime observed in the case
$\rho\ll 1$ where the acceptance rate is large is very similar to that
observed in models with a continuous breaking of the replica symmetry,
like the $SK$ model\cite{CuKuRi} or finite-dimensional
spin-glasses\cite{Rieger,Rieger2} where no evidence of plateau is found
in the region of values of $t_w$ explored.

\subsubsection{The case $\rho > 1$}

In this case $Erf(\alpha)\sim \frac{e^{-\al^2}}{\al}$ since $\alpha$ is
large, i.e. the acceptance is always very small.  In this regime the
rescaled time constant eq.(\ref{renorm}) is again small because
$Erf(\al)$ is small (it decreases with $\rho$ like
$\rho e^{-\rho^2}$). For values of $t_w\ll \frac{exp(\rho^2)}{\rho}$ the
system is in the short-time regime where a plateau is present and aging
is present (i.e. the $C_0(t_w,t_w+t)$ strongly depends on $t_w$).  This
is clearly appreciated in figure 3 where we show the $C_0(t_w,t_w+t)$
for different values of $t_w=1,10,100,1000$ at $T=0.4$, $\rho=5$ and the
numerical solution of eq.(\ref{Apb4}). In the regime $t_w\ll
\frac{exp(\rho^2)}{\rho}$ there is a plateau which
increases with $t_w$ and eventually converges to the value $q_{EA}=1-T$
for times $t_w$ of order $\frac{exp(\rho^2)}{\rho}$. This is trasient 
regime where the decay to the plateau does not satisfy time-translational
invariance. No equivalent regime is found in the Langevin dynamics of the 
spherical Sherrington-Kirkpatrick model.

The behavior shown in figure 3 is very similar to that found in systems
with strong freezing in the low temperature regime where the acceptance
rate is quite low. Generally these are systems with one step of replica
symmetry breaking like the spherical or Ising $p$-spin model
($p>2$)\cite{CHS,Gardner}, the Potts Glass model with $p$-states
($p>4$)\cite{PG}, the ROM (random ortoghonal model)\cite{ROM} and
frustrated systems without quenched disorder like the Bernasconi
model\cite{BERNA} and other type of models \cite{others}.  All these
systems are characterized by the presence of a dynamics with very low
acceptance rate and the existence of a dynamical transition different of
the statical one. While the statics of the aforementioned class of
models is much different to that of the spherical SK model 
the dynamics shows interesting similarities.


\subsection{Zero-temperature dynamics}

In this subsection we are interested in extracting the large-time
behavior of the dynamical equations at zero temperature. There is no
reason in the zero-temperature case to get the same large-time behavior
as in the Langevin case, the time-rescaling eq.(\ref{renorm})
being ill defined.  Because in the zero-temperature regime the
acceptance rate goes monotonically to zero (i.e. the parameter $\al$
diverges) the coefficients $a(t), b(t), c(x,t)$
in eq.(\ref{coefene},\ref{eqcx}) have a non trivial large-time limit. Then
we expect a dynamical regime much different from that predicted in the
Langevin dynamics.  In fact, we will see that there is no limiting case
in which the Langevin dynamics is found.

At $T=0$ the coefficients $a(t),b(t), c(x,t)$ of
eqs.(\ref{coefene}),(\ref{eqcx}) become

\be
\begin{array}{lr}
a_{T=0}(t)=&-\frac{2\,\al\,e^{-\al^2}}{E\sqrt{\pi}}\\
b_{T=0}(t)=&-(\frac{\rho^2\,Erf(\alpha)}{2}+\frac{4\al\,e^{-\al^2}}
{\sqrt{\pi}})\\
c_{T=0}(x,t)=&\frac{1}{2}\rho^2\,\ll e^{x\la}\gg \,Erf(\al)
\end{array}
\label{coefT0}
\ee

Also the equation (\ref{eqa301}) for the acceptance rate reads,

\be
A(t)=\frac{Erf(\alpha)}{2}
\label{eqa304}
\ee

Note that the large time dynamics is governed by the
time-dependent parameter $\alpha$ which diverges in the infinite-time
limit. Unfortunately the dependence of the coefficients on $\al$ is
strongly non-linear, which makes the mathematical treatment of the
dynamical equations highly non trivial. However we can derive a nonlinear
equation for the energy alone by means of the adiabatic approximation 
explained in the next subsection. Although a rigorous derivation of the
adiabatic approximation is not known to us, the resulting equation yields
an excelent approximation to the large-time behavior as given by the
numerical solutions of the exact dynamical equations.

We can analyze qualitatively how the system evolves at zero temperature.
Suppose the system starts from a random initial configuration $\s_i=\pm
1$ such that $E(t=0)=0$ and $B_1(t=0)=1$. The energy monotonically
decreases to the ground state energy $E=-\frac{J_{max}}{2}=-1$ while
$B_1$ decreases also to zero. In the large time limit $\alpha$ diverges
and the acceptance rate goes to zero. There are two different regimes in
the dynamics. The first one is an initial regime where $\alpha$ is small
and the acceptance rate is nearly $1/2$. This corresponds to a gaussian
$P(\De E)$ (eq.(\ref{eqa38})) with width $\rho \sqrt{B_1}$ larger than
the position of its center ($\rho^2E$). In this case, the changes of
configuration which increase or decrease the energy have the same
probability. The energy decreases fast in this regime because the
acceptance is large.  The second regime appears when $B_1$ is so small
in order that $\alpha$ becomes large. In this case the acceptance is
very small (it goes like $\frac{exp(-\alpha^2)}{\al}$) and the dynamics
is strongly slowed down. The system goes very slowly to the
equilibrium. It is in this second regime where the adiabatic
approximation developed in the next subsection applies.

The adiabatic approximation shows that the
large-time behavior in the Monte Carlo case is quite different form the
Langevin case. This is not a surprise. Even for the simple
harmonic oscillator, one can show that at zero temperature the dynamics
is different in the MC and in the Langevin case
\cite{PaRi}. This is related to the non ergodic nature of the dynamics
at zero-temperature in which only the changes which
decrease the energy are accepted. This is clearly shown in figures
4, 5, 6, 7. In figure 4 and 5 we show the acceptation rate and energy 
as a function of time for three different values of $\rho$ by
numerically solving the off-equilibrium equations eq.(\ref{Apb3}) with
the coefficients (\ref{coefT0}). 

We note two important results. First, as previously said, we observe
that $\al$ is the parameter which governs the dynamics.  More concretely
figure 4 shows that the magnitude of $\al$ separates two different
regimes. In the regime $\al\ll 1$ the acceptance rate is $0.5$ while it
falls down rapidly to zero in the regime $\al\gg 1$.  Second, the large
time behavior of the energy shown in figure 5 is strikingly different
from that expected in the Langevin dynamics at zero temperature where
$E(t)+1\simeq O(1/t)$ \cite{CugDean}. The energy decays like $1/log(t)$
(plus corrections) as shown below.

Figure 6 shows the $C_0(t_w,t_w+t)$ as a function of $t$ for four
different values of $t_w$ by numerically solving the off-equilibrium
equations eq.(\ref{Apb4}) with the coefficients (\ref{coefT0}). Note
that the large $t$ behavior of the $C(t_w,t_w+t)$ is strikingly
different in the MC and the Langevin case since no trace of the power
law decay $t^{-\frac{3}{4}}$ is observed and the $\frac{t}{t_{w}}$
scaley behaviour is lost. The correlation function
strongly freezes (i.e. remains very close to one) for values of $t_w$
such that the acceptation rate at that time has already jumped to zero
(see figure 4). This is reasonable because a very low
number of accepted changes implies a very small change of the
correlation function.

\subsection{The adiabatic approximation}

We now discuss the adiabatic approximation 
which gives the correct asymptotic large time behavior of the energy
and the acceptation rate at zero temperature.

The equation of the energy eq.(\ref{eneTf}) can be written as a
function only of the parameter $\al$, 

\be
\frac{1}{E}\frac{\partial E}{\partial t}=-\frac{1}{2}\rho^2
Erf(\al)\,+\,\frac{\rho^2 e^{-\al^2}}{2\al\sqrt{\pi}}~~~~~~.
\label{Apc1}
\ee

This is an exact equation for the energy. Unfortunately we cannot solve
it because we do not know the time evolution of $\al$. Once $\al$ is
known we also know the time evolution of the energy, hence also all the
moments $h_k$.  In this sense, the parameter $\al$, and consequently the
acceptance rate, fully determine the dynamics. In the large-time regime,
where $\al$ is very large, we can expand the error function $Erf(\al)\simeq
\frac{e^{-\al^2}}{\sqrt{\pi}\al} (1-\frac{1}{2\al^2})$ and we get the
simple equation,

\be
\frac{1}{E}\frac{\partial E}{\partial t}=\frac{\rho^2
e^{-\al^2}}{2\sqrt{\pi}\al^3} 
\label{Apc2}
\ee

We solve this equation using an adiabatic approximation which turns out
to be the correct large-time solution as one can check numerically
solving the equation (\ref{Apb3}). We replace $\al$ by an effective
parameter which depends on off-equilibrium quantities via a
quasi-equilibrium relation. In our case we use the quasi-equilibrium
relation eq.(\ref{alpha}) with $\beta=\frac{1}{T}=\frac{1}{\mu^*(E+1))}$
where $\mu^*$ is a renormalized parameter which has the physical meaning
of an effective temperature. Note that $\mu^*=2$ in the previous equation
yields the equilibrium relation $E=-1+\frac{T}{2}$ in the low
temperature phase. Substituting in (\ref{Apc2}) and using $B_1=-2\mu^*
E(E+1)$ we obtain, in the large-time limit (i.e. $E \simeq -1$) and 
neglecting higher orders in $E+1$,

\be
\frac{\partial E}{\partial t}=-\frac{2}{\rho\sqrt{\pi}} 
(\mu^* (E+1))^{\frac{3}{2}}\exp(-\frac{\rho^2}{4\mu^*(E+1)})
\label{Apc3}
\ee

In principle, $\mu^*$ is an unknown parameter. In the simplest adiabatic
approximation $\mu^*=2$. This corresponds to equipartitioning in the
surface of constant energy where the equilibrium relation $B_1=-4E(E+1)$
is fulfilled at all times. It is possible to show \cite{PaRi} that for
$\mu^*=2$ the equation (\ref{Apc3}) is the equation of the energy for a
simple harmonic oscillator with Hamiltonian ${\cal
H}=\frac{1}{2}m\omega^2 x^2$ where $m\omega^2=\la^{max}=2$. Physically
this means that the system relaxes as a simple harmonic oscillator with
a value of $m\omega^2$ determined by the maximum eigenvalue of the
spectrum. Physically a value of $\mu^*<2$ means that the system relaxes as
an effective single harmonic oscillator with effective mass
$m\omega^2=\frac{4}{\mu^*}$ larger than the single oscillator mass
$m\omega^2=2$ \cite{PaRi}.


The large-time solution of eq.(\ref{Apc3}) can be easily worked out. One
finds that the parameter $\al$ diverges like $(\log(t))^{\frac{1}{2}}$,
the energy decays like $E(t)=-1+O(\frac{1}{log(t)})$ and the acceptation
rate goes like $A(t)\simeq \frac{1}{t}$. All these quantities have
non-trivial subdominant corrections. In figure 7 we show the effective
parameter $\mu^*=-\frac{B_1}{2E(E+1)}$ as a function of time obtained by
solving eq.(\ref{Apb3}) for different values of $\rho$. Note that the
effective parameter $\mu^*$ converges to a time-independent value in the
asymptotic large time limit. By numerically solving the adiabatic
equation (\ref{Apc3}) we have checked that the relaxation of the energy
is in excellent agreement with the numerical solution of
eq.(\ref{Apb3}). This is a check of the adiabatic approximation which
yields the values $\mu^*\simeq 0.087,0.72,1.65$ for $\rho=0.1,1,5$
respectively. Note that these quoted values of $\mu^*$ are smaller than
$2$.  

\section{Conclusions}

In this work we have presented the analytical solution of the Monte
Carlo dynamics of the spherical Sherrington-Kirkpatrick model. This is a
very simple spin-glass model where full computations can be carried
out. In order to analytically solve the MC dynamics we have used the
technique of the generating function. The generating function allows us to 
close the dynamical equations in a similar way to how it has been done in other 
glassy models whithout quenched disorder. The present work is a step towards
the use of this powerful technique in the spin glass context. As an
example, we have solved the Langevin dynamics of the spherical SK
model. In the context of the MC dynamics, the generating 
function has been used to derive closed expressions for one-time quantities, 
like the generalized moments $h_k$, and for two-time quantities such as the 
correlation or response functions. The formalism of the generating function
may be used to compare the results obtained with both dynamics, MC and Langevin.

By solving such a simple model, we lose the subtleties of having a replica 
broken low-temperature phase where the presence
of a large number of metastable states makes dynamics much more interesting
\cite{Cri}. Nevertheless our results for the MC dynamics make it clear how 
relevant for off-equilibrium dynamics is the acceptance rate, a quantity 
which is absent in Glauber or Langevin dynamics. Since a major part of the 
numerical work on glassy systems employs MC dynamics, its analytical study is 
of the outmost importance to be used as a guide when trying to extract 
conclusions from numerics.

One of the most important results which emerge from this work concerns
the importance of the acceptance rate in the off-equilibrium dynamics.
We have seen that the main differences between the Langevin and the MC
dynamics appear whenever this rate is small and for times not too large.
The Langevin dynamics is obtained in the limit $\rho\to 0$ with an effective 
rescaling of time $t\to t'=t\frac{\rho^2\beta}{2}$ and the acceptance ratio is
always $1$ (cf.\ equations (\ref{coefene}) and (\ref{eqcx})). For 
finite $\rho$ and finite temperatures we have
seen that the Langevin regime is found for times $t\simeq\rho^{-2}$ when
$\rho$ is small and for times $t\simeq \exp(\rho^2)/\rho$ when $\rho$ is
large. In the intermediate regimes very many different
behaviors are observed, specially for $\rho>1$ where the acceptance is
small. The dynamics observed in figure 3 reminds one a lot of what is observed
in models with one step of breaking (see for instance
\cite{KrMe,MiRi}). Furthermore, at zero temperature, the MC dynamics
yields a completely new behavior where the system rapidly freezes as
soon as the acceptance goes to zero. No analogous regime is found in
the case of Langevin dynamics  where the transition from finite to zero
temperature is smooth.

We would like to comment three different directions which we think it
would be interesting to explore. The first concerns the adiabatic
approximation which is used to obtain closed equations for one-time moments. As 
it has been shown, this approximation suggests
that the system behaves as a system of uncoupled harmonic oscillators
with a renormalized mass larger than the maximum mass corresponding to
the largest eigenvalue of the Wigner spectrum. Although this suggestive
approximation works extremely well, it would be interesting to
understand it better. Future research should deal with a systematic derivation
of the adiabatic approximation for one-time quantities (for instance, understanding
the dependence of the effective parameter $\mu^*$ as a function of $\rho$),
and its extension to the study of the long-time behavior of correlation 
functions. We think that the adiabatic approximation is a consequence of
the relevance of entropic effects for the dynamics of the system. In the
low acceptance regime, the system spends a long time searching
configurations of lower energy in a very inefficient way. In this regime
(not present when the dynamics is Langevin's) the adiabatic approximation works 
very well.  More generally one would like to improve it in
order to find a systematic way to close the dynamical equations without
having to use the full generating function. The work done by Coolen, Sherrington and
collaborators \cite{CS} and that on the Backgammon model \cite{BG} is a
step in this direction.  The second direction to explore is the study of models 
with one step of replica symmetry breaking (like the $p$-spin spherical
spin-glass model \cite{CHS}) using the technique of the generating
function.  This is an interesting problem whose solution would allow us to 
obtain closed equations for one-time quantities like the energy and probably a
set of higher moments. At present it is not clear whether this is
possible or whether we can only obtain equations which relate the correlation 
and the response function \cite{CHS,CuKu}.  Finally we want to
mention that the type of differential equations studied here, (i.e. semilinear
evolution equations like Eq.\ (\ref{eqb38}) whose coefficients depend 
self-consistently on the first moments of the solution), appear often in the 
study of the dynamics of quite different models in statistical physics. 
Let us cite among others the problems of synchronization of populations of
coupled oscillators \cite{kuramoto}, arrays of Josephson junctions
\cite{swift}, and plasmas or self-gravitating systems \cite{Padman,BCS}.
Often these problems can be described by non-linear causal equations where the 
time evolution of an observable at time $t$ depends on all its previous history
plus self-consistent relations for functions appearing in the equations. 
A general mathematical study of these types of equations would be welcome.

\section{Acknowledgments}

The work of L.L.B., F.G.P and F.R. has been supported by the DGICYT
of Spain under grant PB92-0248 and the EC Human
Capital and Mobility Programme contract ERBCHRXCT930413.
We are grateful to Silvio Franz for a careful reading of the manuscript, and to
Conrado Perez-Vicente and Alex Arenas for stimulating
discussions on this and related subjects.

\appendix
\section{The equation of motion}

In this section we write the general equation of motion associated to 
the general joint probability distribution 

\be
P(\De E,\De O)=K\,\exp(-\frac{(\De E-a)^2}{2b^2})\,
\exp(-\frac{(\De O-c)^2}{2d^2})
\label{Apa1}
\ee

\noindent
where $c=e+f(\De E-a)$ and $a,b,c,d,e,f$ are in general time-dependent
parameters, $\De E$ stands for the energy change and $\De O$ the change
of any observable (generalized moment, correlation or response
function). In the MC dynamics $a=-\rho^2\,E$, $b^2=\rho^2 B_1$
(with $B_1$ given in eq.(\ref{eqa39})) and $c,d,e,f$ depending on the
particular observable. The constant $K=(4\pi b^2 d^2)^{-\frac{1}{2}}$
normalizes the probability distribution.

The equation of motion for the  observable $O$ is,

\be
\frac{\partial O}{\partial t}= \int_{-\infty}^{\infty}dx\,\int_{-\infty}^{\infty}dy\,y P(x,y)\,w(x)
\label{Apa2}
\ee

\noindent
where $w(x)=Min(1,e^{-\beta x})$ is the Boltzmann
factor. Straightforward computations yield 

\be
\frac{\partial O}{\partial t}=\frac{1}{2}\Bigl (e\,Erf(\alpha)\,+\,
(e-f\beta b^2)\,\exp(-a\beta+\frac{b^2\beta^2}{2})\,
Erf(\sqrt{\frac{b^2\beta^2}{2}}-\alpha)\bigr )
\label{Apa3}
\ee

\noindent
where $\alpha=-\frac{a}{(2b^2)^{\frac{1}{2}}}$ and coincides with
eq.(\ref{eqa303}) with the previously quoted values of $a, b$. 

At zero temperature the equation of motion is, 

\be
\frac{\partial O}{\partial t}=\frac{1}{2}\Bigl (e\,Erf(\alpha)\,
-f\sqrt{\frac{2b^2}{\pi}}\,\exp(-\al^2)\Bigr )
\label{Apa4}
\ee

\noindent
with the same previous definition of the parameter $\al$.

\section{Solution of the differential equation}

In this section we present the general solution to the generating
function $g(x,t)$ in eq.(\ref{eqb38}) using
the method of characteristics,

\be
\frac{\partial g(x,t)}{\partial t}=a(t)\frac{\partial g}
{\partial x}\,+\,b(t) g\,+\,c(x,t)
\label{Apb1}
\ee

In order to solve this equation we make the following change of
variables $x\to u(t)=x+\int_0^t\,a(t')dt'$ and eq.(\ref{Apb1}) becomes 

\be
\frac{d \hat{g}(u,t)}{d t}=b(t) \hat{g}(u,t)\,+\,c(u,t)
\label{Apb2}
\ee

\noindent
which is a linear differential equation easily solvable. The final
result is,

\be
g(x,t)=g_0(x+\int_0^t\,a(t')dt')B(t)+B(t)
\int_0^tdt' \frac{c(x+\int_{t'}^t\,a(t'')dt'',t')}{B(t')}
\label{Apb3}
\ee

\noindent
where $B(t)=\exp(\int_{0}^t\,b(t')dt')$ and $g_0(x)=g(x,t=0)$ is the
initial condition. Once the $g(x,t)$ has been obtained one can also get
the generating function of the two time quantities. For instance the
$K(x,t',t)$ associated to the correlation function in eq.(\ref{eqc35})
is given by,

\be
K(x,t',t)=g(x+\frac{1}{2}\int_{t'}^t\,a(t'')dt'',t')
\Bigl (\frac{B(t)}{B(t')}\Bigr)^{\frac{1}{2}}~~~~~~.
\label{Apb4}
\ee

\vfill\eject

\vfill
\newpage
{\bf Figure Captions}
\begin{itemize}

\item[Fig.~1] Relaxation of the energy $E(t)-E^{eq}$ as a function of
time for $\rho=0.1$ and temperatures $T=0.2,0.4,0.6$. The points are the
MC simulations data ($N=500$, one sample), the lines are the
analytical solution of the MC dynamical equations and the dashed line is
the large time behavior described by the Langevin dynamics rescaling
the time as described in the text.

\item[Fig.~2] $C_0(t_w,t_w+t)$ with $\rho=0.1$, $T=0.4$ for
$t_w=1,10,100,1000$. The points are the MC simulations data ($N=500$,
one sample), the lines are the analytical solution of the MC dynamical
equations and the dashed line is the large time behavior described by
the Langevin dynamics rescaling the time as described in the text.

\item[Fig.~3] $C_0(t_w,t_w+t)$ with $\rho=5$, $T=0.4$ for
$t_w=1,10,100$. The points are the MC simulations data ($N=500$, data
averaged over 5 samples) and the lines are the analytical solution of
the MC dynamical equations.

\item[Fig.~4] Acceptance rate $A(t)$ at $T=0$ as a function of time for
$\rho=0.1,1,5$ obtained by solving the MC dynamical equations (lines)
compared with MC simulations (points) ($N=500$, one sample).

\item[Fig.~5] Energy $E(t)+1$ at $T=0$ as a function of time for
$\rho=0.1,1,5$ obtained solving the MC dynamical equations (lines)
compared with MC simulations (points) ($N=500$, one sample).

\item[Fig.~6]  $C_0(t_w,t_w+t)$ at $T=0$ with $\rho=0.1$ for
$t_w=1,10,100,1000$ (from bottom to top) obtained solving the MC dynamical i
equations. 

\item[Fig.~7] Ratio $-B_1/(2E(E+1))$ as a function of time at $T=0$ for
different values of $\rho$. The large-time limit yields the
effective parameter $\mu^*$.

\end{itemize}

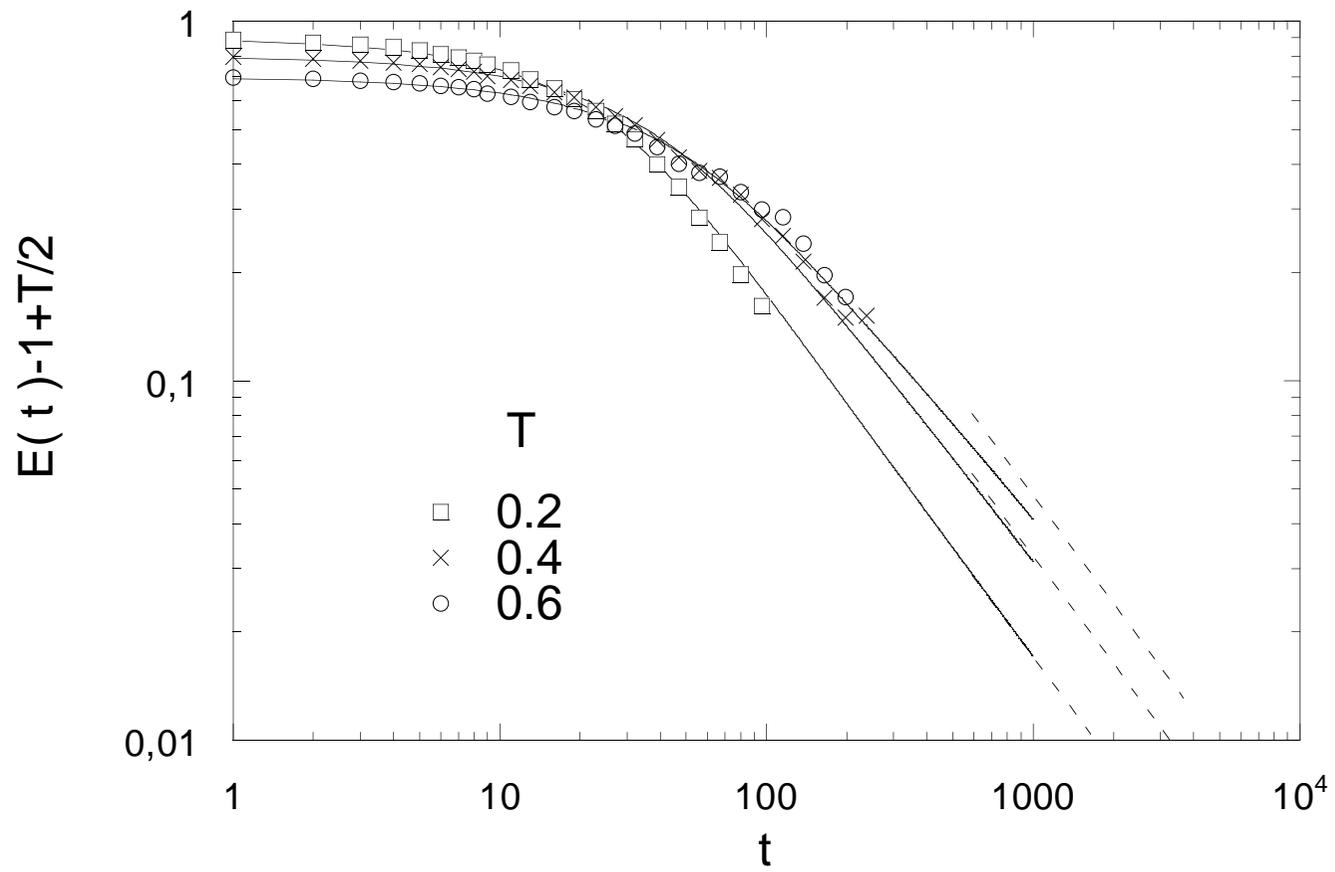
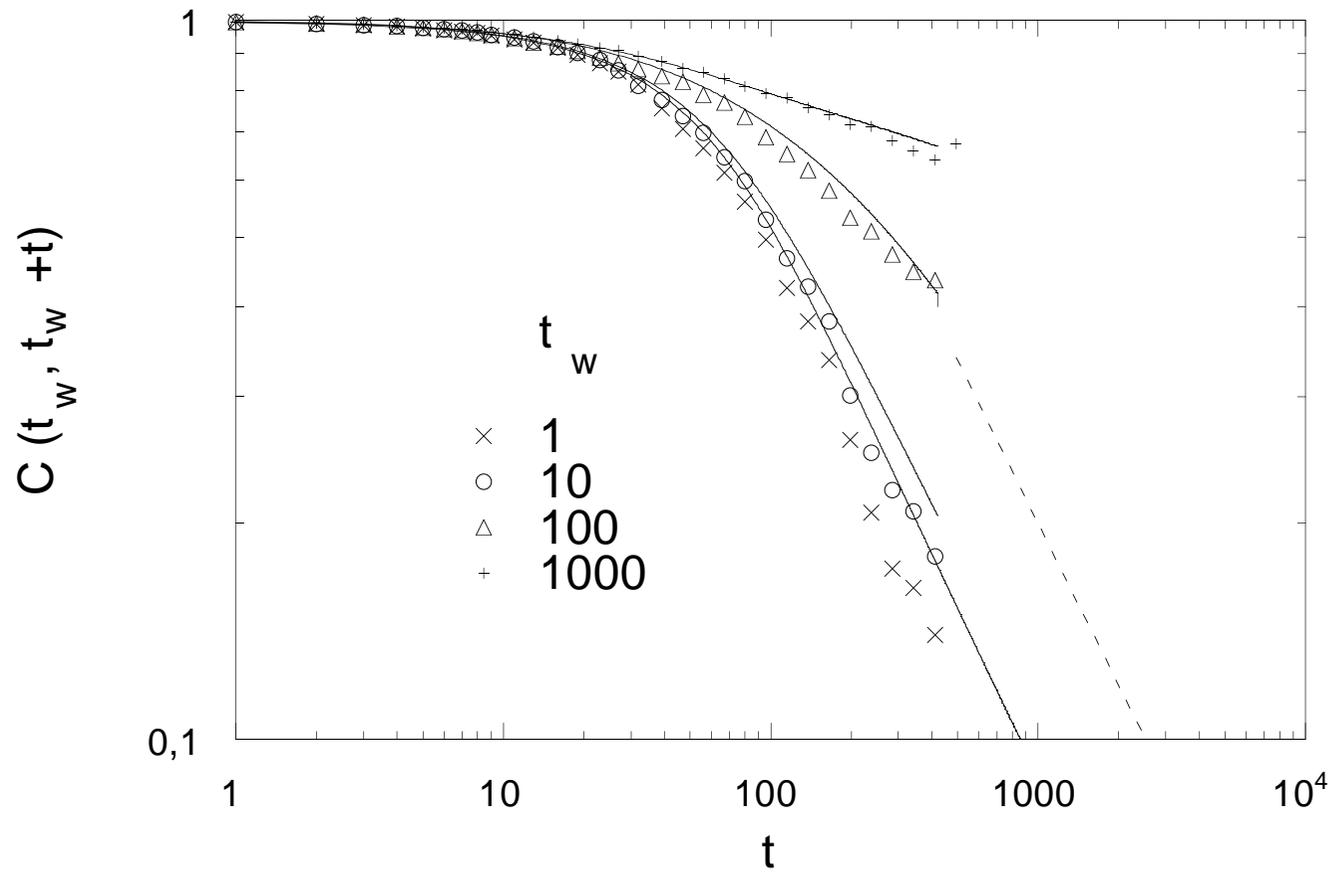
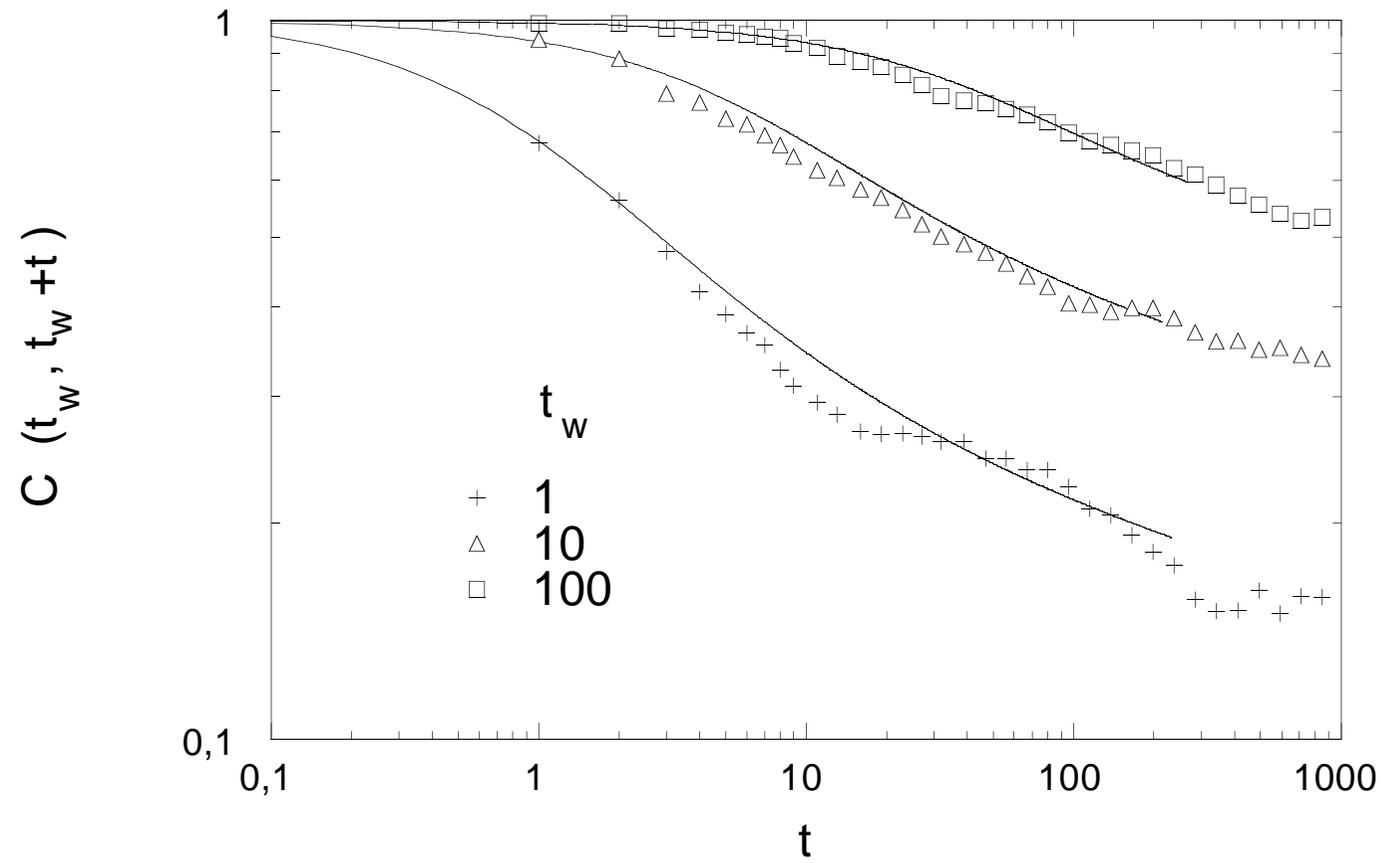
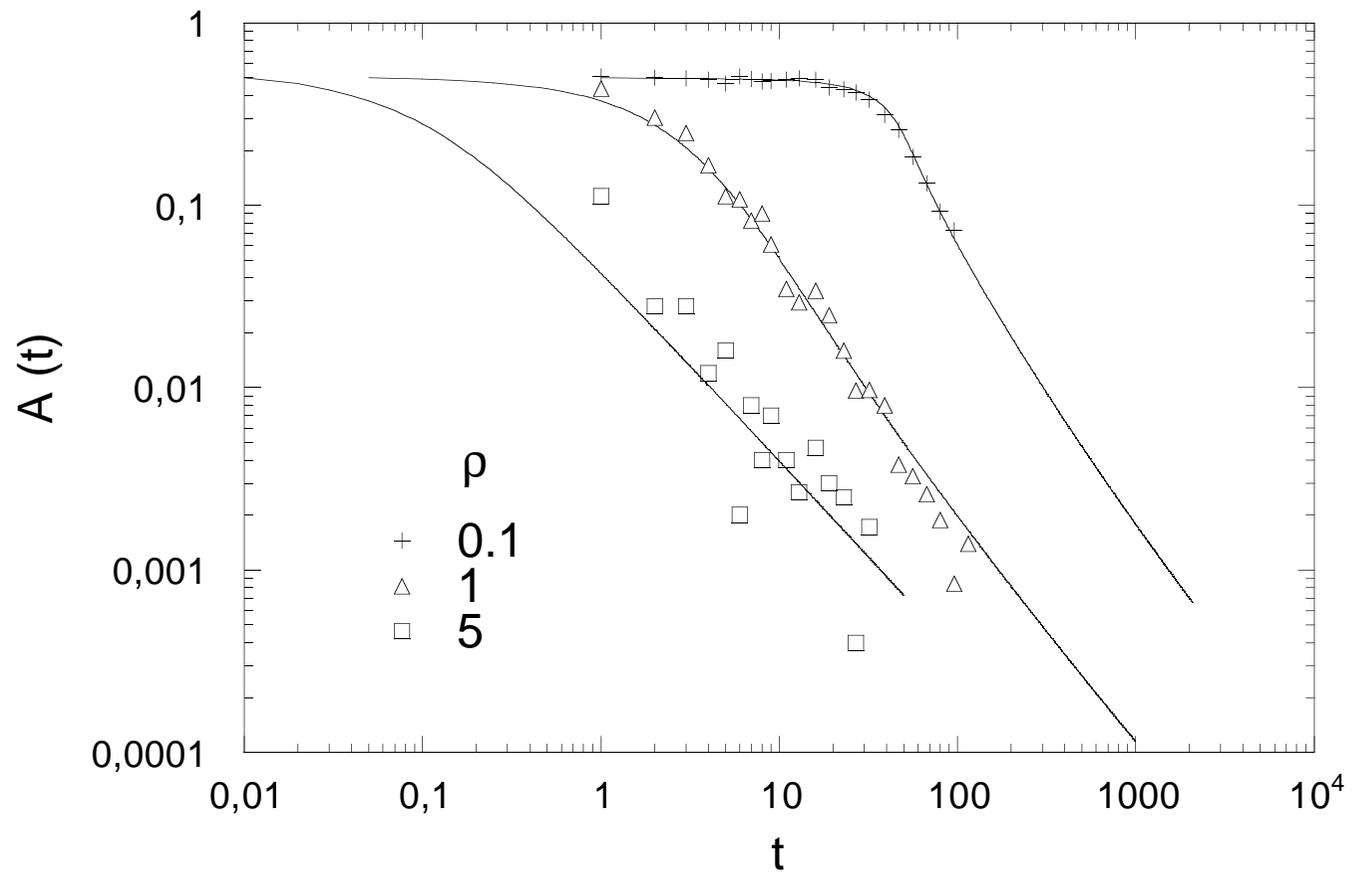
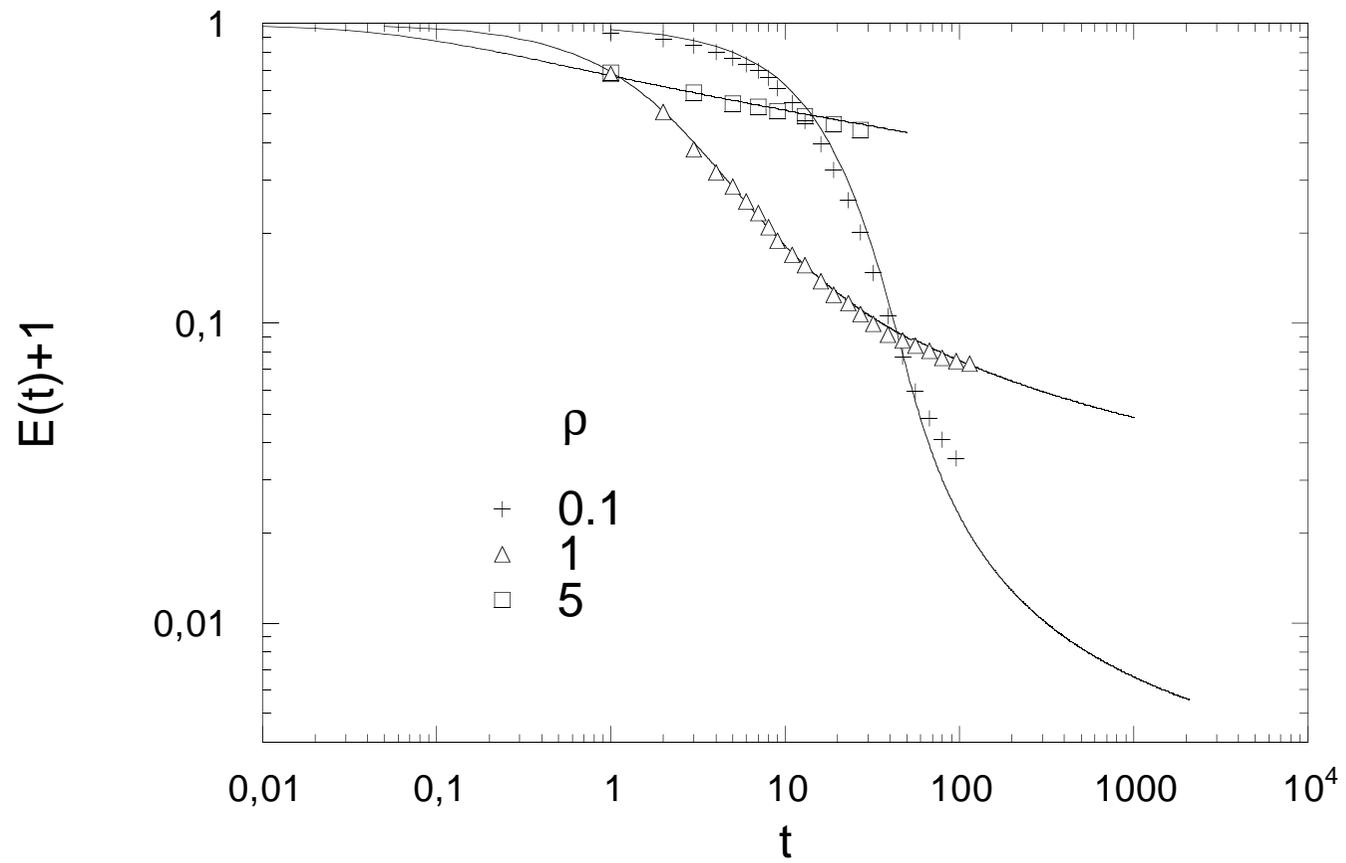
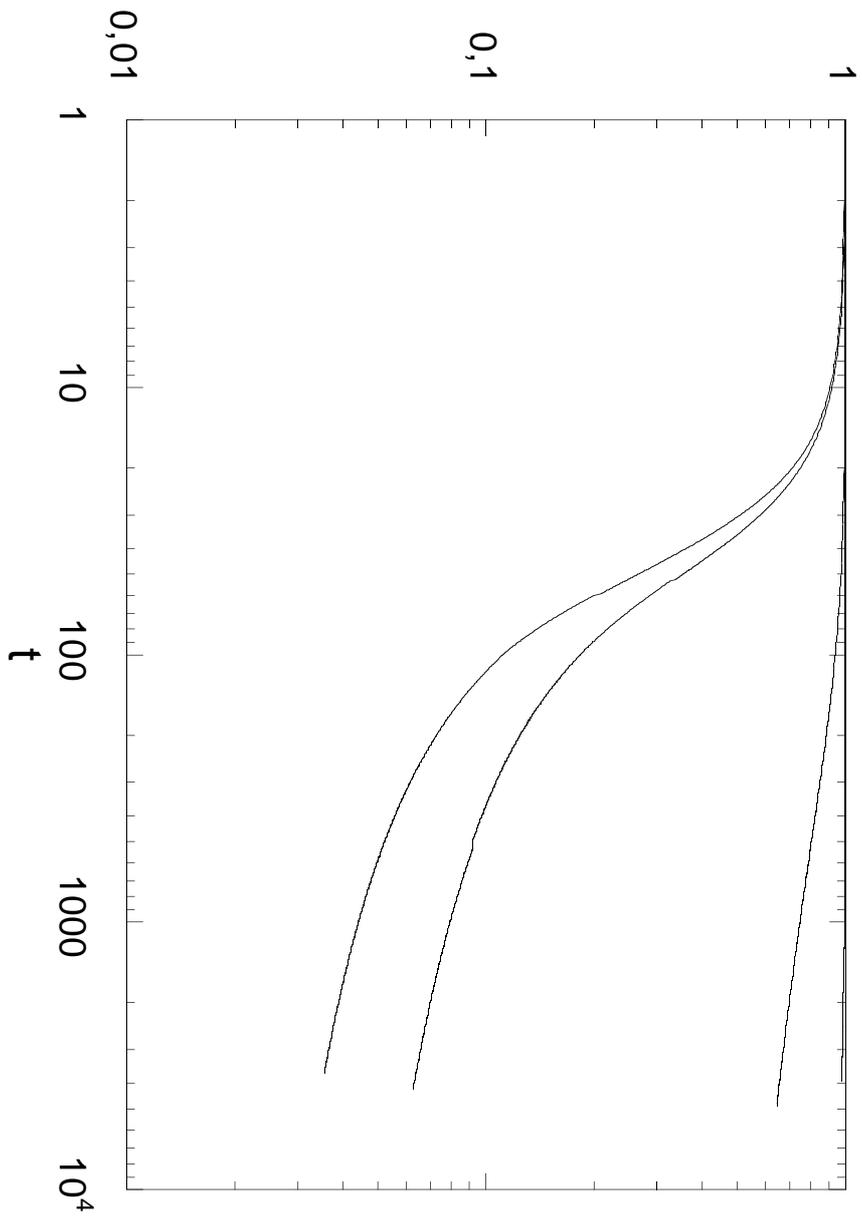
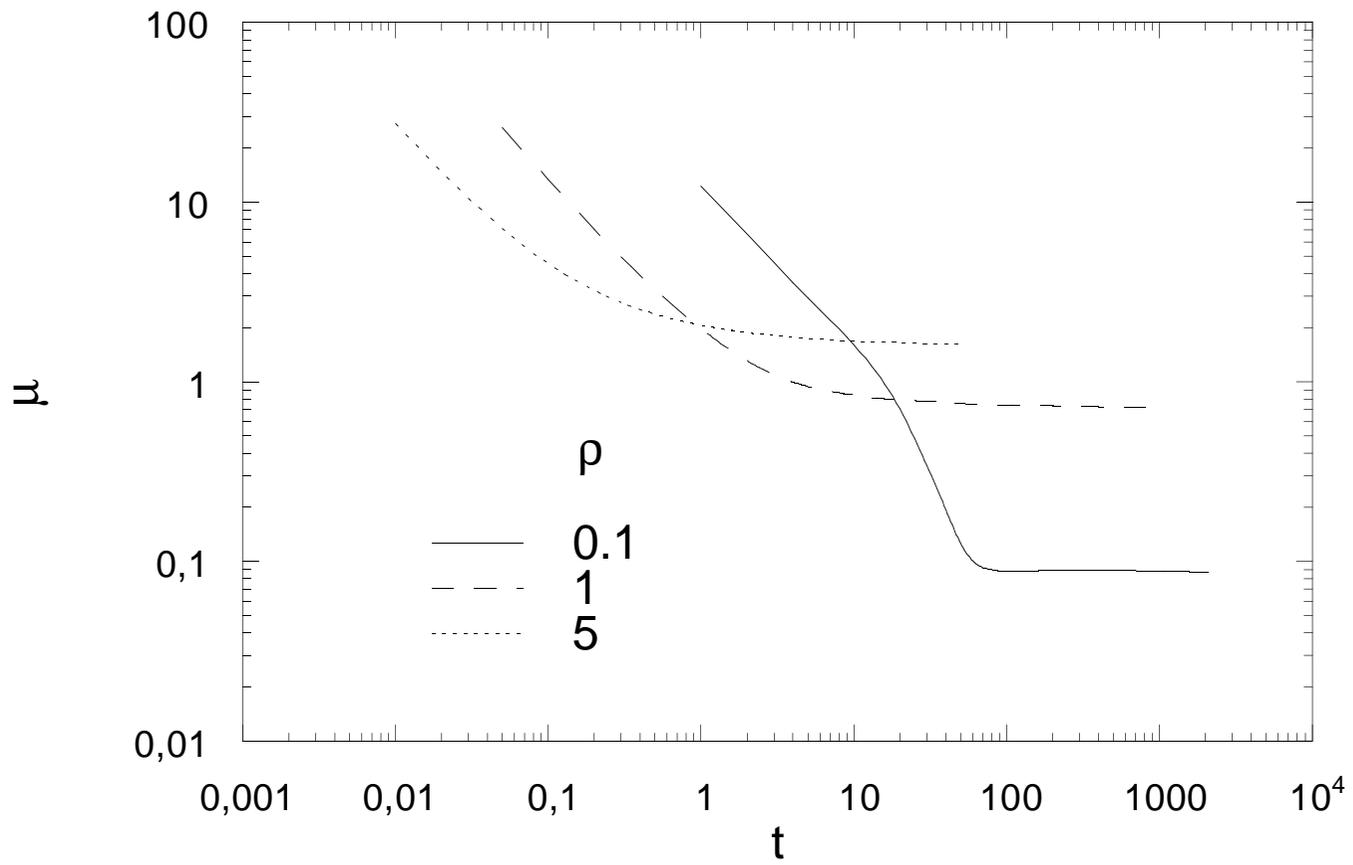
\end{document}